\documentclass[10pt,letterpaper]{article}
\usepackage[top=0.85in,left=2.75in,footskip=0.75in]{geometry}

\usepackage{changepage}

\usepackage[utf8]{inputenc}

\usepackage{textcomp,marvosym}

\usepackage{fixltx2e}

\usepackage{amsmath,amssymb}

\usepackage{cite}

\usepackage{nameref,hyperref}

\usepackage{microtype}
\DisableLigatures[f]{encoding = *, family = * }

\usepackage{rotating}

\usepackage{color}
\usepackage[table]{xcolor}

\usepackage{setspace} 
\usepackage[aboveskip=1pt,labelfont=bf,labelsep=period,justification=raggedright,singlelinecheck=off]{caption}

\makeatletter
\renewcommand{\@biblabel}[1]{\quad#1.}
\makeatother

\date{}

\usepackage{lastpage,fancyhdr,graphicx}
\usepackage{epstopdf}

\begin{document}

\newgeometry{left=2cm,top=2cm,bottom=2cm}

\begin{flushleft}
{\Large
\textbf{\newline{Predicting the Impact of Electric Field Stimulation in a Detailed Computational Model of Cortical Tissue}}\newline
}

Frances Hutchings\textsuperscript{1,*},
Christopher Thornton\textsuperscript{1},
Chencheng Zhang\textsuperscript{2},
Yujiang Wang\textsuperscript{1},
Marcus Kaiser\textsuperscript{1,2},
\\
\bigskip
\bf{1} Interdisciplinary Computing and Complex BioSystems, School of Computing Science, Newcastle University, UK\\
\bf{2} Department of Functional Neurosurgery, Ruijin Hospital, Shanghai Jiao Tong University, China
\\
\bigskip


%
%





* frances.hutchings@newcastle.ac.uk

\end{flushleft}
\pagebreak
\section*{Abstract}

\paragraph{Background}
Neurostimulation using weak electric fields has generated excitement in recent years due to its potential as a medical intervention. However, study of this stimulation modality has been hampered by inconsistent results and large variability within and between studies. In order to begin addressing this variability we need to properly characterise the impact of the current on the underlying neuron populations.  

\paragraph{Objective}
To develop and test a computational model capable of capturing the impact of electric field stimulation on networks of neurons. 
\paragraph{Methods}
We construct a cortical tissue model with distinct layers and explicit neuron morphologies. We then apply a model of electrical stimulation and carry out multiple test case simulations. 
\paragraph{Results}
The cortical slice model is compared to experimental literature and shown to capture the main features of the electrophysiological response to stimulation. Namely, the model showed 1) a similar level of depolarisation in individual pyramidal neurons, 2) acceleration of intrinsic oscillations, and 3) retention of the spatial profile of oscillations in different layers. We then apply alternative electric fields to demonstrate how the model can capture differences in neuronal responses to the electric field. We demonstrate that the tissue response is dependent on layer depth, the angle of the apical dendrite relative to the field, and stimulation strength.
\paragraph{Conclusions}
We present publicly available computational modelling software that predicts the neuron network population response to electric field stimulation.

\section*{Author Summary}

Low intensity electric field stimulation of the brain has been proposed as a promising therapy for a number of neurological disorders. However, there are issues with large variability in the reported success of this stimulation approach. In order to explore the root of this variability, we present a novel, freely available modelling tool that can predict how any given electric field will change the activity of a population of neurons. 
Our model includes different neuron types and layers, with simplified morphologies used to model individual neurons. This allows us to simulate a large number of connected neurons in a model of brain tissue. We can then simulate an electric field across the tissue. To demonstrate its use, we show how the software replicates published findings and can make new predictions. In particular, we find that the brain tissue’s response to stimulation is dependent on the layer depth and the angle of the neurons in relation to the stimulating electric field.


\section*{Introduction}

\noindent
Transcranial direct current stimulation (tDCS) is a form of non-invasive brain stimulation which involves applying electric fields to the brain via two or more electrodes on the scalp. Characteristically, the current that passes through the skull to the brain is not strong enough to elicit action potentials in affected neurons, but has been shown to modulate neuronal excitability with a knock-on effect on firing likelihood \cite{Reato2013}. 

One potential application of this technology is alleviating symptoms of neurological disorders, however, despite promising preliminary studies there is still debate over whether tDCS is an effective intervention at all for brain disorders. The main reason for scepticism stems from the large variability in reported results. Meta reviews have concluded that whilst there is some evidence of efficacy for treatment of major depression \cite{Ferrucci2009,Shiozawa2014}, chronic pain  \cite{Vaseghi2014} and stroke recovery \cite{Butler2013}, these results are weakened by variability both within and between studies \cite{Brunoni2016,Shiozawa2014,Butler2013,OConnell2014}.

There are a number of explanations for this variability, including (but not limited to): 1) variation in individual brain morphologies, 2) differing background brain activity, 3) inconsistency in placement of electrodes and parameters used in treatment \cite{Ramaraju2018}, and 4) differences in the electric fields generated even with the same montage due to physical variations \cite{Laakso2019}. Additionally, the majority of studies use small numbers of participants rendering statistical significance difficult to obtain. 

Predictive models may be one way of unpicking the influences of the above variables \cite{Wang2015}. Currently, the majority of computational models for tDCS are finite element models (FEMs) which predict the current density in the target region. These FEMs can anticipate the locations in the brain that stimulation from an arbitrary electrode montage will reach and can predict differences in electric fields based on brain morphology \cite{Neuling2012, Datta2013}. However, these models are unable to predict the response of the underlying tissue to the stimulation. The model we introduce is intended to fill this gap. We do this by applying FEM derived electric fields to a spiking neural network model, replicating key features of the cortical tissue response to stimulation. 

Existing models of neural responses to stimulation take one of two approaches. 1: Tissue level, where stimulation is modelled as a direct, predetermined depolarisation in the neuron membrane potential \cite{Molaee2013,Reato2010,Dutta2013}. 2: Single neuron level, where a finite element model field is applied and the neuron membrane changes are calculated \cite{Rahman2013,Arlotti2012}. We combine both approaches, and by using simplified neuron morphologies we are able to simulate large networks of neurons while applying stimulation without a predetermined neuron response. In this way the model can provide a prediction of the tissue-level effects of stimulation from individualised FEM current densities.

In order to test the validity of the model we replicated an experimental study \textit{in silico}, where the authors applied a 4mV/mm DC electric field to an \textit{in vitro} ferret visual cortex slice \cite{Frohlich2010}. The authors reported three main findings which we also observe in the model: 1) small net depolarisation in single neurons, 2) acceleration of intrinsic slow oscillations, and 3) spatial dynamics across layers are modulated but still maintained. Finally, as a model prediction, we examine the response of the model to other field strengths and orientations.

\noindent
We replicated the experimental conditions from an existing \textit{in vitro} study in VERTEX to demonstrate that the model of electrical stimulation can also display the main reported effects of stimulation. We present a cortical model replicating slow oscillations ($\sim$1Hz) in the ferret (\textit{Mustela putorius furo}) visual cortex, as in the \textit{in vitro} study. We show that our model concurs with the main observations of a previously published experimental study. We then vary the orientation of the stimulation field to predict how the changes affect the network activity. 

\subsection*{Validation with an experimental electrical stimulation study}

\begin{figure}[h]
 \begin{center}
 \includegraphics[scale=1.35]{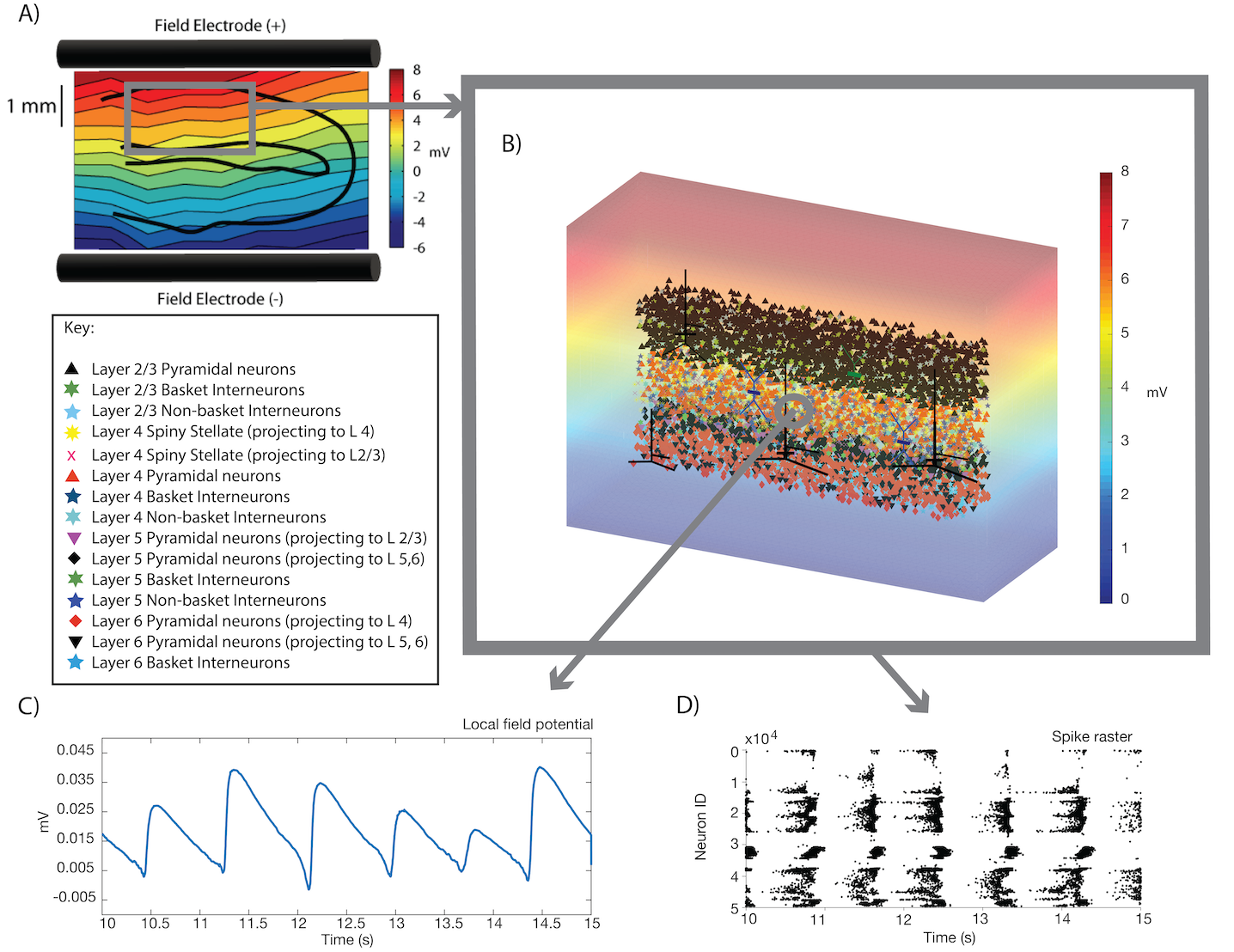}
 \end{center}
 \caption{{\bf Slow oscillations alongside a schematic of the cortical slice in an electric field.} Part A shows an image from the Fr\"{o}hlich paper of the electric field that was used in their experimental set up \cite{Frohlich2010} with the gray box indicating roughly the equivalent area that the VERTEX slice is simulating. Part B shows the ferret slice neurons, indicated by shapes for different neuron types which are identified in the key below, within a DC electric field with a 4mV centre point similar to the electric field used in the Fr\"{o}hlich experiments. Recording electrodes are shown as black circles and were evenly spaced across the tissue. Part C shows the activity traces in an average Local Field Potential plot, and part D shows this in a spike raster plot, both illustrate slow oscillations which are close to 1Hz in frequency.}
\label{fig:sliceschematic}
\end{figure}

\noindent
Fr\"{o}hlich \textit{et al} \cite{Frohlich2010} demonstrated the impact of electric current stimulation on ferret visual cortex exhibiting slow ($<1$Hz) oscillations. Part A) of Fig.~\ref{fig:sliceschematic} shows the field used in the Fr\"{o}hlich experiments \cite{Frohlich2010}. For comparison, Fig.~\ref{fig:sliceschematic} B) shows an example of an electric field generated in the VERTEX stimulation framework with the soma positions of different neuron types plotted inside the field. The slice schematic additionally shows an overview of the neuron types and soma positions in each layer in the ferret visual cortex slice model. Fig.~\ref{fig:sliceschematic} C) and D) show slow oscillations simulated in our model with an average of the local field potential across all recording electrodes and a spike raster plot. 

\subsubsection*{Small net depolarisation of individual neurons}

\begin{figure}[h]
 \begin{center}
 \includegraphics[scale=0.75]{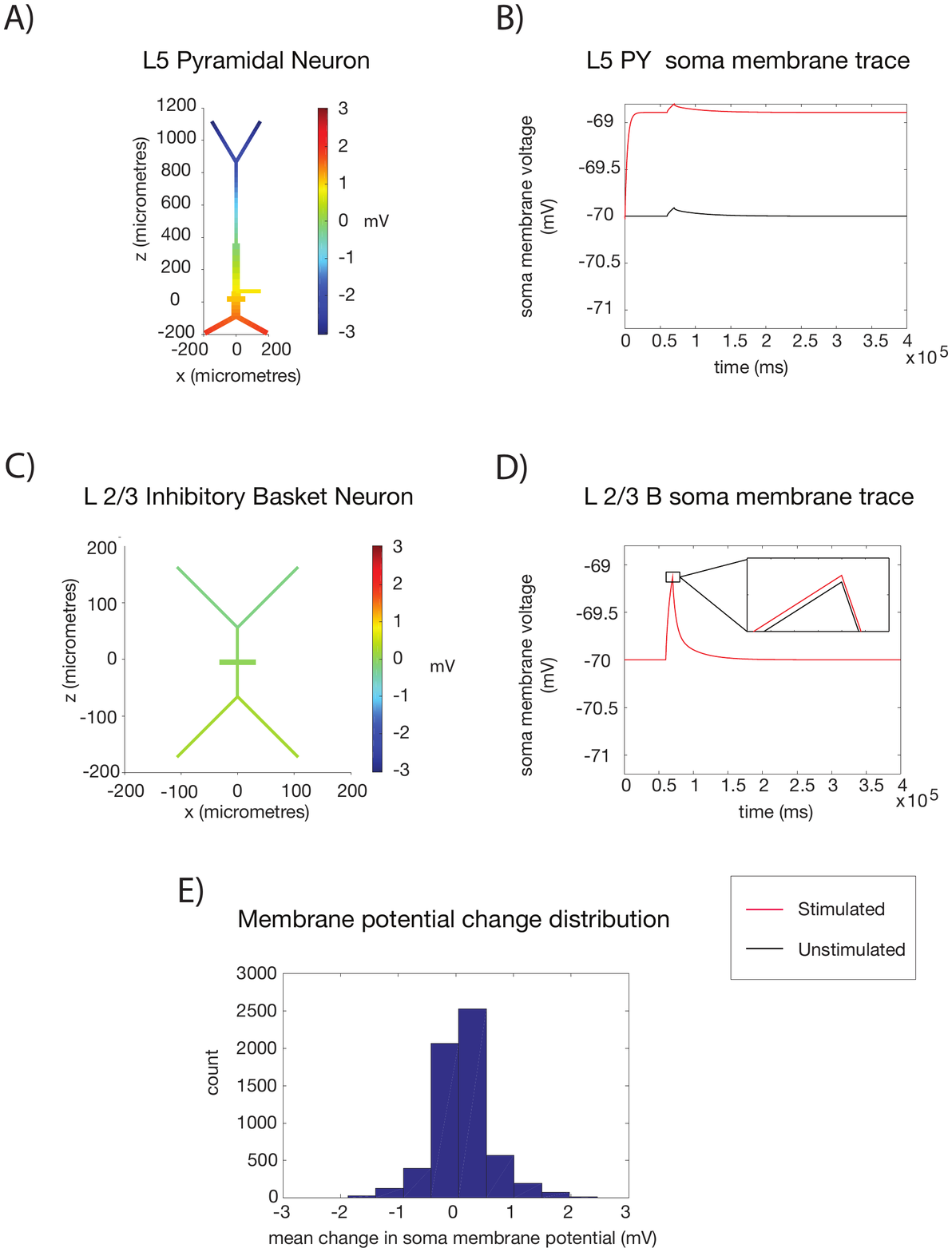}
 \end{center}

\caption{\textbf{Figure 2. Neuron membrane potential changes in the presence of an electric field.} Plot A shows the response of an individual layer 5 pyramidal neuron, using the reduced morphology present in the VERTEX simulator, to the electric field. Compartments are coloured according to the difference in membrane potential in mV. The field was applied with a strength of 4mV/mm in the absence of noise. Part B shows the corresponding soma membrane potential trace over time comparing the no stimulation with the stimulation condition. Again there is no noise present, although a brief step current is applied to induce a spike of activity. These plots are echoed below in parts C and D, where the simulations are repeated for the inhibitory basket interneuron. The colour scales show that the change in neuron compartment membrane potentials are much smaller for inhibitory interneurons compared to pyramidal neurons. This is shown all the more clearly in part D where the changes in soma membrane potential over time shows a very minimal shift. The magnified inset shows membrane potential differences. Part E shows the average change in soma membrane potential for a subset of the neurons across the model, recorded during the slow oscillation experiments.}
\label{fig:vmchange}
\end{figure}

\noindent
The Fr\"{o}hlich study took intracellular recordings of infragranular (layer 5/6) neurons, finding a net depolarisation effect in the presence of the 4mV/mm electric field of 1.29 mV, $\pm$ 0.2 mV. This depolarisation is within the range found in Fig.~\ref{fig:vmchange} for the layer 5 (infragranular) pyramidal neurons, both in the soma changes shown in part A and in the soma membrane change in part B. The computational model allows us to predict that this level of depolarisation is the case for pyramidal neurons, but not for other neuron types - the interneuron stimulated in parts C and D of Fig.~\ref{fig:vmchange} shows a much smaller change in membrane potential. Part E of Fig.~\ref{fig:vmchange} gives an idea of the variability in these results when background noise and slow oscillations are present, showing soma membrane potential changes for a subset of neurons across the slice model. There is an overall positive skew to the membrane potential shifts, such that the small subthreshold change in somatic membrane potential is replicated in the VERTEX model.

\subsubsection*{Acceleration of intrinsic oscillations}

\noindent
Acceleration of the slow oscillations in the presence of a DC electric field is another experimental result from Fr\"{o}hlich \textit{et al}. Fig. 3 (A) shows the field induced acceleration of the oscillation as measured through the multi unit activity and recorded in vitro by Fr\"{o}hlich \textit{et al.} \cite{Frohlich2010}. We show how our model replicates this in Fig. 3 (B) with a corresponding plot of spike counts over time. This is an example from a single simulation, replicating the pattern found in the multiunit activity of the experiment. Part C shows the change in frequency over all simulations, and whilst it is not so obvious that there is an acceleration in every case, there is a clear heavy tail to the distribution in favour of acceleration. The mean supports this, lying above zero. Part D shows the percentage changes in spike counts due to stimulation, split by neuron type. In this plot there is a general positive shift, particularly for the excitatory pyramidal neuron populations, which could be an explanation for the acceleration in the majority of cases.

\subsubsection*{Modulating, but not overriding, intrinsic spatial dynamics}

\noindent
A third observation from the Fr\"{o}hlich paper is that the spatio-temporal profile of the local field potential across layers is maintained with stimulation. This was reported from multisite electrode recordings spanning the depth of the slice. They observed a modulation similar to the overall pattern of increasing oscillation frequency, however this does not change the existing pattern of layer differences.

Fig.~\ref{fig:slowlayers} shows LFP measures split for different layers to illustrate this. Part A shows the averaged LFP for different layers, with the red trace corresponding to the stimulated conditions and the black traces corresponding to the no stimulation condition. The form of the oscillations expectably change throughout the layers, most notably with an inversion of the oscillations in layer 6 compared to the upper layers. However, as in the experimental results, this layer dependent oscillation form is retained with stimulation. 

While layer patterns are retained, there are notable differences due to stimulation illustrated in plots B-D of Fig.~\ref{fig:slowlayers}. Part B shows the amplitude shifts across layers, with a general decrease in amplitude evident in across layers. Part C shows a more distinct pattern down the layers for the DC shift, with a reversal in sign between layers 2-3 and layer 5. The frequency shifts in part D reveal a more subtle change. The distributions in the top and bottom layers appear to have a heavier tail towards positive frequency shifts, indicating accelerations with mean values above zero. The frequency shift appears more dependent on the system noise than the DC shift or amplitude. 



\subsection*{Model based prediction: the role of stimulation field angle and neuron geometry}

\noindent
The VERTEX simulator can capture differences in the response to varied stimulation field orientations and strengths, both at the network level and in individual neurons. 
We examine the impact of field orientation in a version of the model without noise present in order to ensure that the changes observed can be independent of the neuronal firing activity in the model. 

Fig.~\ref{fig:lfpmod} shows the DC shift in LFP for the cortical slice over different field orientations and intensities. The simulations were conducted with no noise is present --- the neurons will therefore plateau to a resting soma membrane potential a short time after initialisation of the model. The lack of noise means that the changes shown are due entirely to the stimulation fields, providing a baseline for the direction/intensity effect. This modulation is plotted for the field angle in degrees, and field intensity (mV/mm). Whilst increasing intensity predictably increases the modulation effect of the stimulation, that modulation is dependent on the angle in relation to the underlying neurons. Therefore just knowing the stimulation intensity is not enough to accurately predict the network behaviour, and this pattern will likely be complicated by the intrinsic noise in the system as well. Looking at the aggregated averaged LFP signal in this instance gives an overview of the changes that can be seen at the population level. These changes likely reflect the neuron level modulation as well as being more clearly influential on the cognitive level effects reported by tDCS experiments. For example, a field that runs diagonal (45$^\circ$ or 135$^\circ$) through the tissue needs to be twice as strong (4mV vs. 2mV) to have the same DC shift effect as a field perpendicular to the tissue (0$^\circ$ or 180$^\circ$).


\section*{Discussion}

\noindent
The VERTEX model is the first to allow the simulation of large populations of individually modelled neurons which can also capture the difference that orientation and morphology makes to the global response to a field. We address why a model with these capabilities is needed and how it fits into current literature, followed by discussion of the extent to which the model captures the features of tDCS applied to an \textit{in vitro} brain slice. 

\subsection*{Comparison with current literature}

\noindent
Studies by \cite{Merlet2013} and \cite{Kunze2016} combined population models of tDCS stimulation with the FEM approach, where MRI derived head models are used to predict realistic electric fields.  We use a detailed model of neuronal tissue rather than a mean field approach, including neuron orientations and simplified morphologies. Point neuron models or mean field approaches cannot account for the impact of electric field orientation, which single neuron models have indicated is important in accurately predicting the neuronal response to stimulation \cite{Rahman2013}. 

There have been a small number of recent studies combining the FEM approach with detailed neuron models \cite{Seo2019,Aberra2018a,Aberra2018b}. Our model also differs from the models of \cite{Seo2019,Aberra2018a,Aberra2018b} as we simulate a large number of interconnected neurons in the field rather than the isolated  neuron morphologies. Networked populations of neurons can produce emergent activity which is hard to predict from studying single neuron models. For this reason, in order to gain a better mechanistic understanding of the electric field stimulation impact on brain tissue it is important to be able to capture stimulation impact on complex, realistic neural networks as well as on single neurons.


\subsection*{Comparison with experimental results}

\begin{figure}[h]
 \begin{center}
 \includegraphics[scale=0.45]{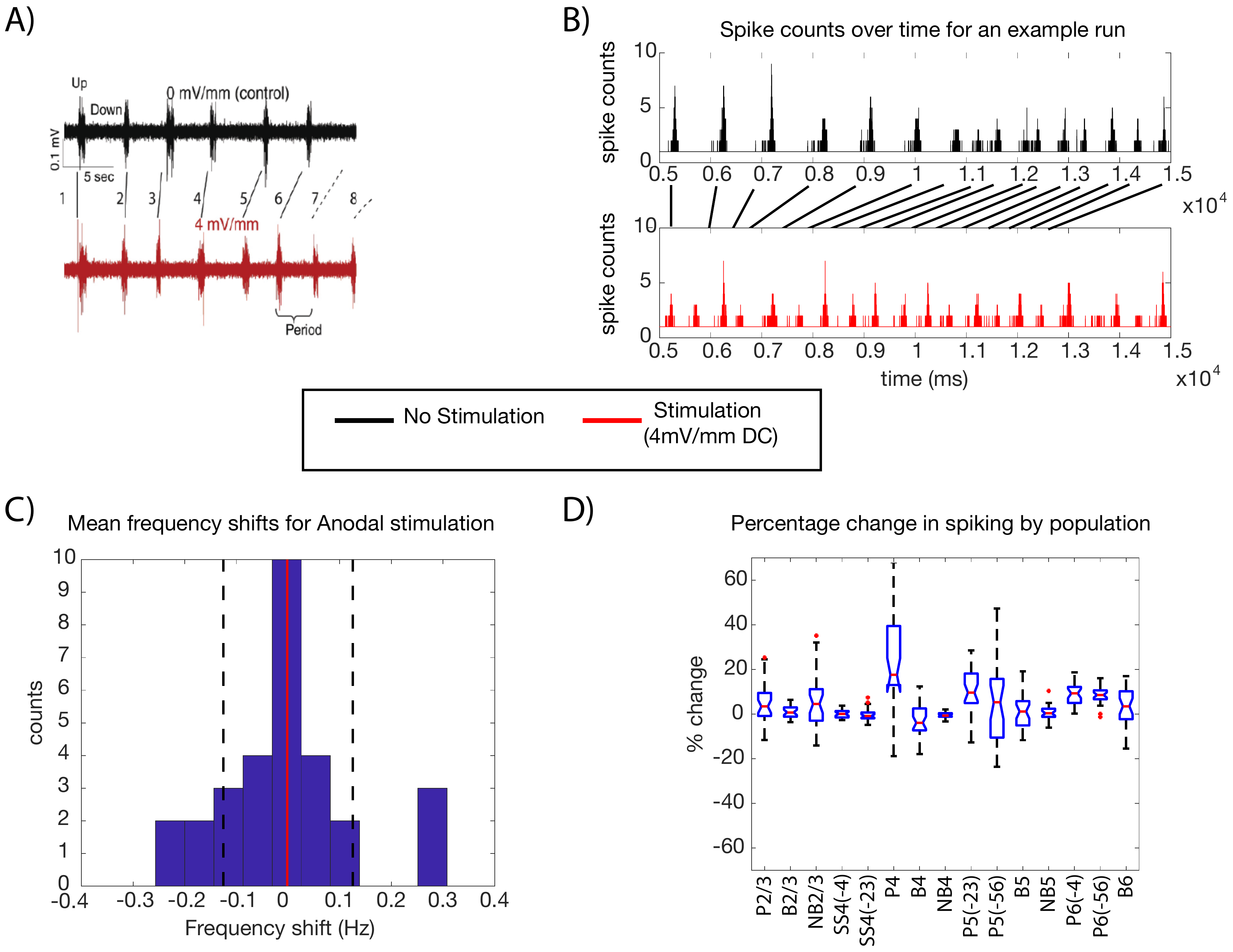}
 \end{center}
 \caption{\textbf{Direct current stimulation applied to the ferret slice.} The plots shown all compare modelling results for tissue stimulated with a DC field of 4mV/mm with the same tissue without stimulation. Part A is one of the original figures from the \cite{Frohlich2010} paper showing the increase in frequency of oscillations with stimulation that they observed. Part B shows the spike counts over time for all neurons as a comparison with the multiunit activity shown in the \cite{Frohlich2010} figure in part A. The acceleration due to stimulation is highlighted using lines connecting the spiking peaks for both conditions. This was plotted for a representative pair of stimulated and unstimulated simulation runs for one noise seed. The histogram in part C shows the mean frequency shifts for pairs of no stimulation and stimulated runs with the same noise seeds, for 30 different simulation runs. The mean is shown by the solid black line and the standard deviations by the dashed black lines. The plot in D shows the percentage change in the number of neurons spiking between no stimulation and stimulated conditions for 30 different noise seed pairs, divided up by the neuron population in the model. The labels on the x axis represent the populations, where P = pyramidal neurons, B = basket interneurons, NB = non-basket interneurons and SS = spiny stellate neurons. The numbers after the letters represent the layers that these neuron types are in. Where there are repeats with brackets (e.g. P5(-23)) these neuron populations are subdivided by their projections into other layers, the layers they project to are indicated in the brackets. (see Fig. S6 for more details).}
\label{fig:dcvalid}
\end{figure}

\begin{figure}[h]
 \begin{center}
 \includegraphics[scale=0.55]{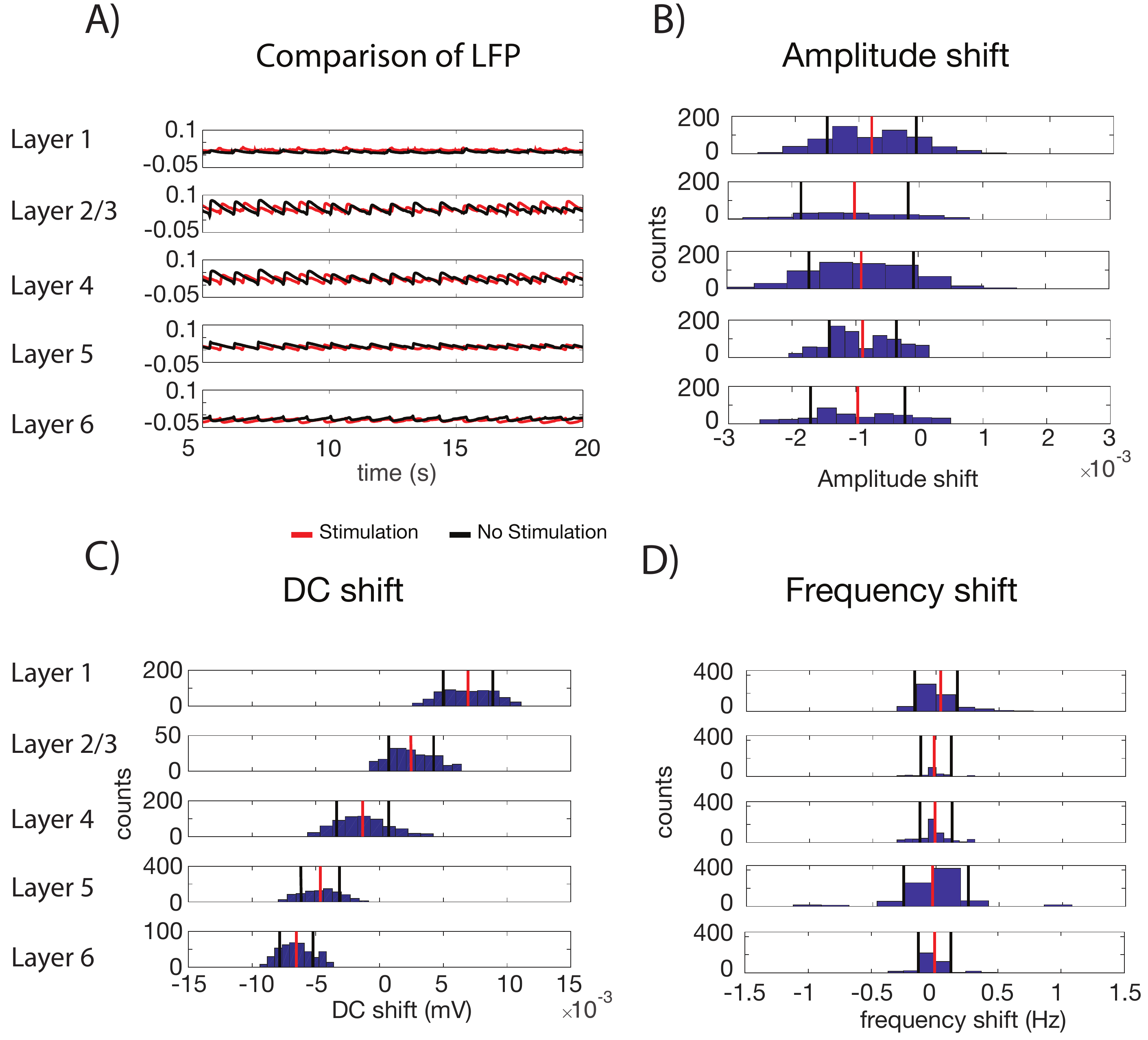}
 \end{center}
 \caption{\textbf{Cortical layer responses to anodal stimulation.} Part A shows the local field potential averaged for each layer for the network with noise present and background slow oscillations, for one noise seed. The black traces are the LFP for the no stimulation condition, and the red traces show the LFP with the DC field applied. The remaining plots show histograms for different measures derived from the LFP and separated by layer. Part B shows the change in amplitude, C the DC shift, D the frequency shift. In each case these are the shifts that occur in the presence of stimulation relative to the no stimulation condition for the same noise seed. The data points in the histograms are for the subset of electrodes present in each layer for 30 different noise seeds. The mean is shown by a red line in each histogram and the standard deviations are shown in black.}
\label{fig:slowlayers}
\end{figure}

\begin{figure}[h]
 \begin{center}
 \includegraphics[scale=0.4]{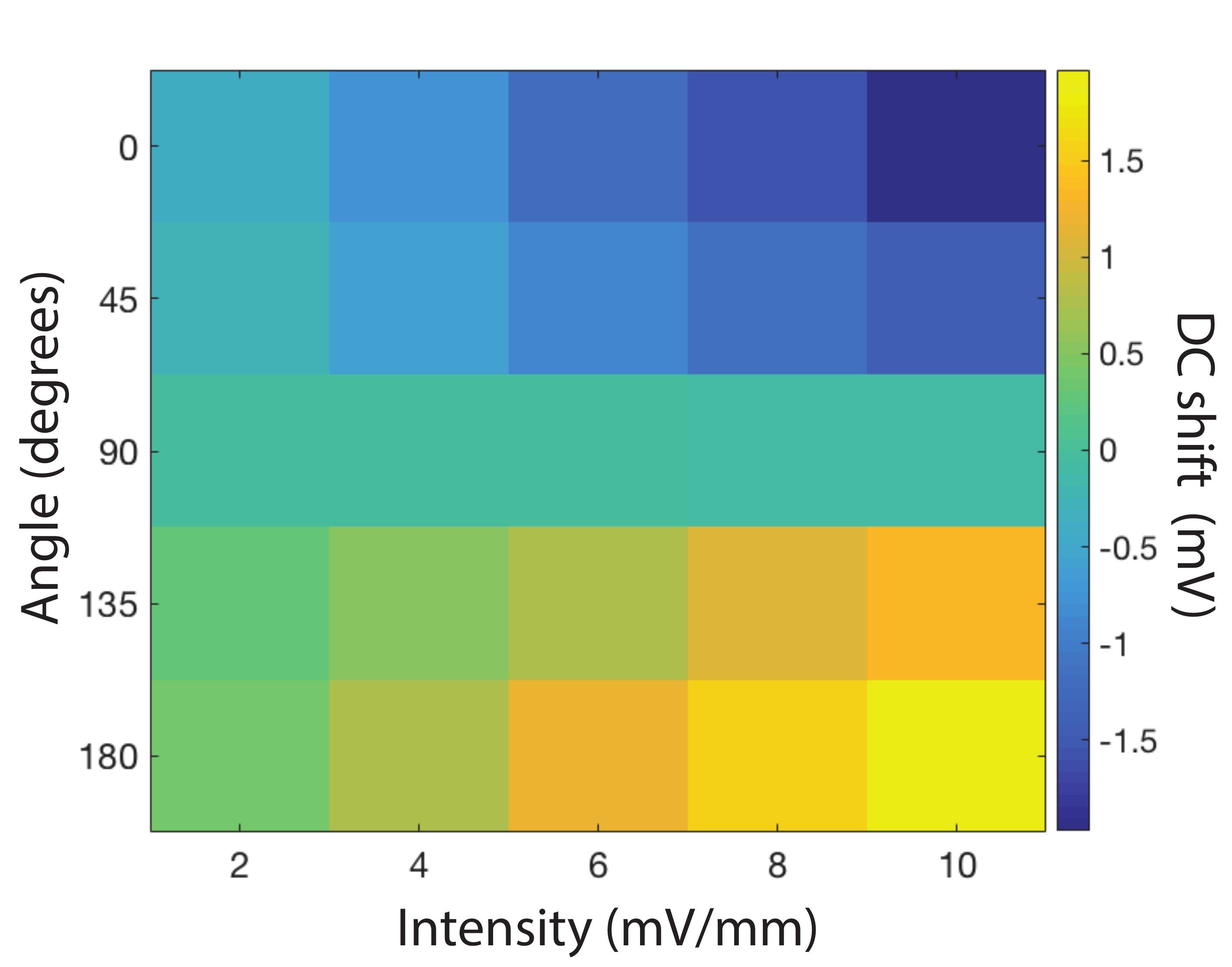}
 \end{center}
 \caption{{\bf Modulation of the local field potential is dependent on both field angle and intensity.} The colours represent the DC shift in the local field potential for the stimulation condition compared to no stimulation, with no noise. The DC shift is plotted for a range of different angles and field strengths to show how the modulation appears to change with these changing parameters.}
\label{fig:lfpmod}
\end{figure}

\noindent
We found that the ferret visual cortex model can replicate major changes observed in tissue behaviour with applied electric fields. However, the effects of the stimulation are not always so clear-cut and noise can overpower some of the effects that are observable in individual cases. This is consistent with the fact that the currents applied in tDCS tend to be subthreshold, meaning the neurons are not induced to spike they are instead just pushed closer to or further from the firing threshold. Given this mechanism it makes sense that the modulation is noise dependent. That being said, for some measures there are clear skews for the majority of simulations, particularly the amplitude and DC shift as shown in Fig.~\ref{fig:dcvalid} parts A and C.



\subsection*{Predictions of layer-wise changes}

\noindent
Figs.~\ref{fig:vmchange} and S4 show an exploration of the impact of the stimulation across different layers, the model easily allows these kinds of results to be presented and analysed. There is no limit to the number of recording electrodes or to the locations they can assume in the model, and there is a certainty as to which layer they occupy. These advantages over experimental recordings allow us to use the VERTEX model to make some predictions about these otherwise hard to measure effects of DC stimulation. 
 

\subsection*{Capturing orientation impact in the cortical tissue model}

\noindent
Fig~\ref{fig:lfpmod} showed that the impact of electric fields cannot be properly understood without taking into account the orientation dependency. This is one reason why the VERTEX model is necessary. Other factors such as neuron density, distribution, and connectivity can also be changed in the VERTEX framework. Given the findings so far we might expect that brain regions with a greater number of large pyramidal neurons would not need as strong a field intensity to get the same effect, so long as the field orientation is aligned with them. However if the orientation is not optimal then almost double the intensity is needed to get the same effect. 

The VERTEX model manages to be computationally tractable for large neuron populations over long run times using the simplified neuron morphologies first presented by Tomsett \cite{Tomsett2015}, and both Fig.~\ref{fig:vmchange} and further work presented in \cite{Thornton2019} show that these morphologies can also capture the impact of an electric field on a given neuron type. Electric fields have the largest impact on longer morphologies orientated in parallel with the field direction, as has previously been reported \cite{Rahman2013}. The fields used in this paper replicate the bar electrode fields applied to \textit{in vitro} slice preparations, as these fields are more appropriate for showing a validation of the presented model. However, it is just as plausible to apply a field more akin to \textit{in vivo} fields used for cognitive experiments or treatment in humans (e.g Fig. S1). 


\subsection*{Limitations of the study}

\noindent
The model presented here is by its nature an approximation of reality, and as such there is a lot of biological detail that is not included, some of which may play a role in the response of networks to stimulation. In the ferret model layers 2 and 3 are combined into one for simplicity, and much of the data used to inform the model was taken from cat visual cortex rather than ferret. While there is evidence that the cat has a similar visual cortex to the ferret \cite{Foxa,Law1988,Rockland1985} it will likely have some differences not captured here. 
Compartments representing the axon are not included in the model morphologies used, and so the propagation of action potentials is not modeled explicitly. Instead we use a delay matrix, which is generated during the initialisation of the network and based on the distance between somas. Another more general limitation of tissue models like the one presented here, is that making predictions about the change in network activity does not easily lead to predictions about the impact at the behavioural level, which particularly for tDCS would be desirable. 

\subsection*{Future work and improvements}

\noindent
The work shown here presents a model that can, in its current form, be applied to mechanistic questions about the impact of electric field stimulation on simulated regions of brain tissue. Upcoming work will be to attempt further experimental validation to lend additional credence to the reliability of this software. Following this, we hope to apply VERTEX to the perhaps more challenging and exciting problem of predicting the optimal stimulation modalities for treatment of patients in clinical applications.

\begin{figure}[h]
 \begin{center}
 \includegraphics[scale=0.2]{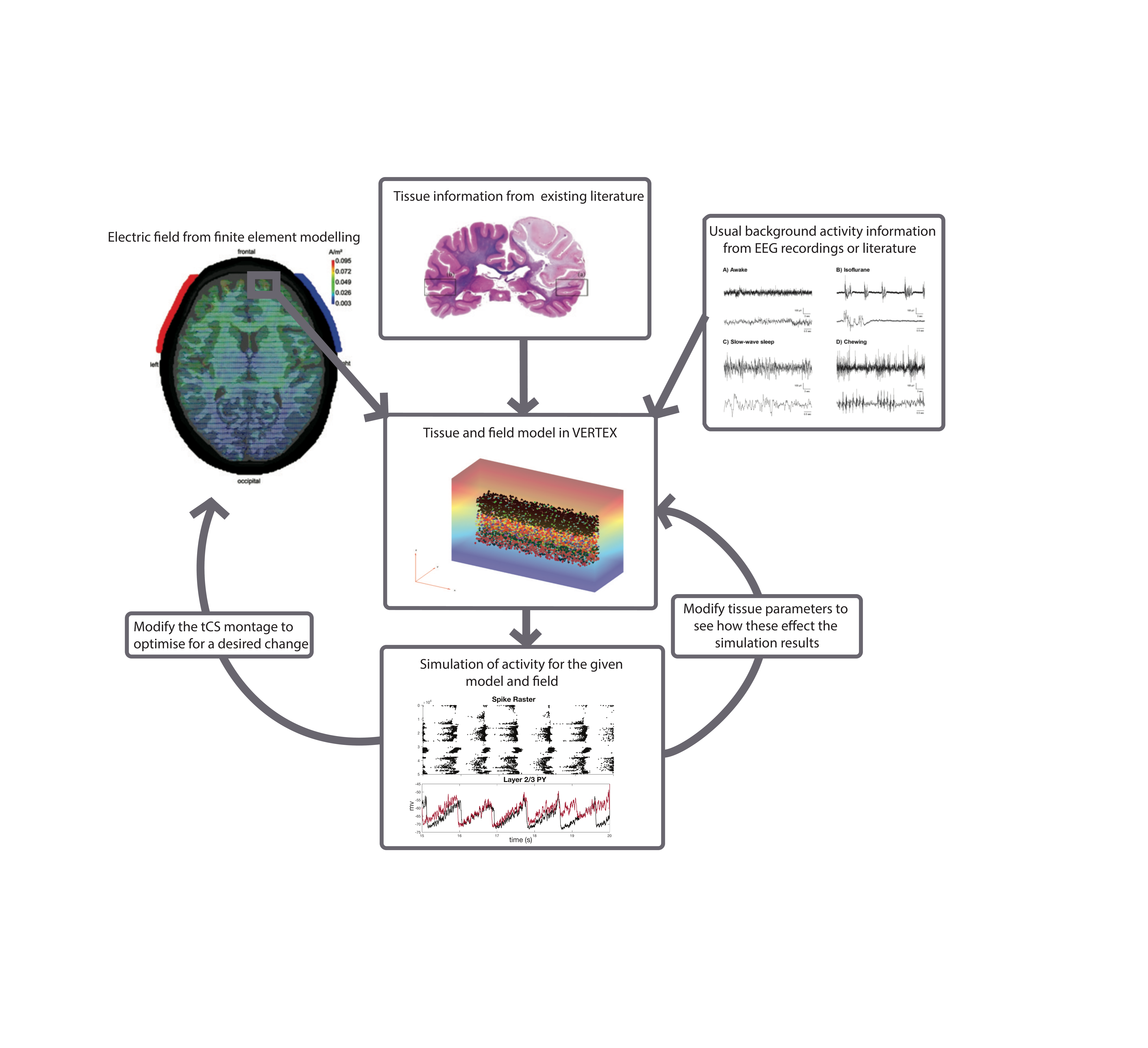}
 \end{center}
 \caption{{\bf Proposed uses of the VERTEX electric field model.} The schematic shown here describes the data that feeds into the model, and the uses of this model for predicting improved stimulation parameters and also for discerning further influences of the electric field impact.}
\label{fig:uses}
\end{figure}

One limitation already noted is that there is still a gulf between predicting neuronal firing effects and actual changes at the level of behaviour. However, by combining the VERTEX model with existing behavioural models it will be possible to speculate about the effects of stimulation and to come up with testable predictions. For example, if we were to assume that anodal tDCS over the dorsolateral prefrontal cortex (DLPFC) would lead to increased cortical excitability and plasticity (as is often the assumption in montages for depression treatment \cite{Brunoni}), then we can test whether a given electric field is liable to create that effect within a model of the DLPFC in the VERTEX simulator. This, or a similar hypothesis, can then be tested under experimental conditions. See Fig.~\ref{fig:uses} for a schematic outline of how this could be accomplished for any given region.

\subsection*{Summary}

\noindent

We have introduced a model of electric field stimulation that builds on the current literature by taking the electric field as an input and calculating the effect on neurons individually. Simplified neuron morphologies mean that large numbers of different neuron types can be simulated simultaneously, including neuron orientation and morphology information. We showed, in agreement with existing literature, that orientation and morphology are particularly relevant for predicting the electric field impact. We demonstrated that the VERTEX model matches existing experimental results, and can allow us to examine the impact of stimulation by layer and neuron type. We present this stimulation model in the hope that it can be used as a tool for 1) better understanding electric neurostimulation, 2) aiding in the choice of stimulation parameters and 3) for predicting success or failure of interventions.

\section*{Methods}

\noindent
The model we introduce is an extension to the VERTEX toolbox, which is built in MATLAB. The finite element modelling required the PDE toolbox and MATLAB version R2016a or later. All simulations in this paper were carried out using MATLAB 2016b. Further details about the VERTEX toolbox can be found at \cite{Tomsett2015}.
The software can be downloaded freely at \url{http://vertexsimulator.org}.

\subsection*{Applying electric fields}

\noindent
We apply field stimulation in VERTEX following the hybrid FEM-cable approach, as outlined by \cite{Joucla2014}, first finding a finite element model solution to the electric field and then applying the result via the cable equation to the neurons themselves. Whilst there are alternative methods, including the whole FEM approach also outlined by \cite{Joucla2014}, the hybrid approach allows for integration with the VERTEX toolbox as well as allowing us to interface VERTEX in future with external FEM software, both commercial, such as HD Explore by Soterix \cite{SoterixMedical2018}, or freely available such as ROAST \cite{Huang2019} or SimNibs \cite{Thielscher2015}.

The finite element models used in this paper were all generated within MATLAB using the partial differential equation (PDE) toolbox. A geometry file was generated in Blender \cite{blender2016}, an open source 3D modelling software, and loaded into MATLAB. The geometry was then imported to the PDE toolbox and boundary conditions were assigned to the faces of the geometry. All boundaries in the geometry which were not acting as current sources were given a Neumann boundary condition, $\sigma \bigtriangledown V \cdot n = 0$. Here, $n$ is the outer vector normal to the boundary, $V$ is the electrical potential in Volts, $\sigma$ is the conductivity and $\bigtriangledown$ represents the gradient. The boundaries representing electrodes or other current sources were given Dirichlet boundary conditions, $h*V=r$. Dirichlet boundaries take two parameters, a weight `$h$', which was always set to 1 for simplicity and an electrical potential `$r$', representing the electrical input into the model in Volts. With $h$ set to 1 this means that $r$ directly represents $V$ in this condition. 

We then solve the Laplace equation  $-\bigtriangledown \cdot (\sigma \bigtriangledown V) = 0 $, following the standard method used in finite element modelling of tDCS \cite{Datta2012,Datta2012,Parazzini2014,Miranda2013}. 
The conductivity $\sigma$ was set to 0.276 S/m \cite{Wagner2004}.
The PDE toolbox inbuilt functions can mesh the given geometry and, once boundary conditions are defined, it then solves the Poisson equation and outputs the electric field and current density for the volume of the given geometry. 

The interaction of the stimulation field with modelled neurons is based on the cable equations described in \cite{Rattay1999}. The electric field values from the finite element model are calculated at the midpoint of each compartment and then the field values are applied to update the axial currents of the neurons. The recent paper by Thornton \textit{et al.} \cite{Thornton2019} goes into more details on the stimulation implementation.

\subsection*{Tracking changes in individual neurons}

\noindent
We set up simulations using individual neurons to show the impact of different electric fields on the simplified neuron morphologies used in the model. Individual neurons were simulated for 400 seconds without noise, and with a small step current applied at the 60 second mark. The stimulation used in all cases was a 4mV/mm uniform DC field.

\subsection*{Cortex model}

\noindent
The cortical slice model was constructed using a modified version of the model presented in \cite{Tomsett2014} which relies heavily on cat data from \cite{Binzegger2004}. The ferret and cat visual cortex have been found to show close similarities \cite{Foxa,Law1988,Rockland1985}, making this a reasonable approximation for replicating the experimental data. The neuron parameters were modified and the slice size, layer boundaries and neuron density (set to 50,000 neurons/${mm}^2$) were all changed to fit the ferret visual cortex. 
The neurons for each layer use the same characteristics and morphologies as the model introduced in \cite{Tomsett2015}. Morphologies are reduced from those presented in \cite{Mainen1996} and compared in \cite{Tomsett2014} against cat visual cortex neurons from \cite{Kisvarday1992,Contreras1997}. 
To produce the slow background oscillation patterns the noise input to different neuron populations was adjusted, as were the synaptic delay parameters. The parameters used are shown in table S9. 
Fig.~\ref{fig:sliceschematic} shows an overview of the cortical slice in an electric field with example recording results exhibiting the slow background oscillations.

\subsection*{LFP recording and analysis}

\noindent
Electrode recording points were specified spanning the whole cortical slice to simulate LFP recordings from multielectrode arrays, using 91 recording electrodes with a 0.1mm spacing.
The LFP is simulated by combining signals from neurons surrounding the recording point as outlined in greater detail in \cite{Tomsett2014}.

We simulated the cortical slice model in the absence of noise for a range of electric field orientations and intensities. To assess the impact of the electric field on on-going oscillatory activity, we ran the cortical slice model for 20 seconds of simulation time and for 30 different noise seeds. The model was initiated with all neurons at their resting potentials, and random noise input was given at each time step to all neuron populations according to the parameters specified in table S9. Electric fields were applied after 100ms until the end of the simulation. In the no stimulation case a zero valued electric field was applied for the same amount of time in order to keep the comparison conditions as close as possible. For each run of simulations (with the different field orientations and the zero-field) the same set of noise seeds were used. This allows for pairs of the same noise seed to be compared, where the differences between the runs are due to stimulation rather than noise. These simulations were repeated for two alternative field orientations. Neuron spikes were recorded by time point and neuron ID for all simulation runs.

We calculated a number of summary measures using the LFP signals averaged within layers. The frequency shift was found using the frequency domain of the signal, Fourier transforming the LFP using inbuilt MATLAB functions and comparing the peak frequency signals. 
Signal amplitude changes were compared by taking the root mean square of the signals after minusing the signal mean to account for the DC shift. 
The DC shift of the signal was approximated by taking the difference between the means of the stimulated condition and the unstimulated condition. These measures were found for each noise seed using the pairs of stimulated and unstimulated conditions which shared the same noise.

\section*{Acknowledgments}
%

\noindent
This work was supported by the CANDO project (http://www.cando.ac.uk/) funded through the Wellcome Trust [102037] and EPSRC [NS/A000026/1] (FH, MK); EPSRC [EP/N031962/1] (MK); CT was supported by a BBSRC DTP MRes/PhD studentship; YW was supported by the Wellcome Trust [208940/Z/17/Z].



\pagebreak


%
%
%


\newgeometry{left=2cm,top=0.2cm,bottom=0.1cm}

 \newpage

\newpage

\newpage

\newpage

\newpage

\restoregeometry

\end{document}